\documentclass[aps,groupedaddress,showpacs,floatfix]{revtex4}
\usepackage{amssymb}
\usepackage{amsmath}
\usepackage{graphicx}
\begin{document}

\title{Replica solution of the Random Energy Model}

\author{Victor Dotsenko}

\affiliation{LPTMC, Universit\'e Paris VI, 75252 Paris, France}

\affiliation{L.D.\ Landau Institute for Theoretical Physics,
   119334 Moscow, Russia}

\date{\today}

\begin{abstract}
The alternative replica technique which involve summation over all integer momenta of the 
partition function and which does not require analytic continuation 
to non-integer values of the replica parameter $n$ is discussed. 
In terms of this technique (which does not involve any replica symmetry breaking
"magic operations") rigorous solution for the average free energy 
of the Random Energy Model is recovered in a very simple way. 
\end{abstract}


\maketitle

\section{Introduction}

In recent years there is a renewed  interest to the mathematical status of the 
replica method widely used in disordered systems during last four decades (see e.g \cite{RSB-general,book}).
For the calculation of thermodynamic quantities averaged over disorder parameters
(e.g. average free energy) the method assumes, first, calculation of the averages
of an integer $n$-th power of the partition function $Z(n)$, and second, analytic continuation of 
this function in the replica parameter $n$ from integer to arbitrary non-integer values
(and in particular, taking the limit $n\to 0$). Usually one is facing difficulties at both
stages of this program. First of all, in  realistic disordered systems
the calculations of the replica partition function $Z(n)$ can be done only using
some kind of approximations, and in this case the status of further analytic continuation 
in the replica parameter $n$ becomes rather indefinite since the terms neglected 
at integer $n$ could become  essential at non-integer $n$ (in particular the limit $n\to 0$)
\cite{zirnbauer1,zirnbauer2}.
On the other hand, even in rare  cases when 
the derivation of the replica partition function $Z(n)$  can be done exactly, 
further analytic continuation to non-integer $n$ appears  to be ambiguous.

The classical example of this situation is provided by the Derrida's Random Energy Model
(REM) \cite{REM}. At present this is one of the best studied models of spin-glasses
(see e.g.\cite{Fyodorov} and references there in) which exhibits non-trivial solution.
 It can be easily shown that in this system the partition function
momenta $Z(n)$ growths as $\exp(n^2)$ at large $n$, and in this case
there are exist many different distributions yielding the same values of $Z(n)$, but 
providing {\it different} values for the average free energy of the system \cite{REM}.
In this situation 
the replica solution which is generally {\it believed} to be correct is obtained via
the "magic operations" of the Parisi replica symmetry breaking (RSB) scheme \cite{RSB-general}
(which in the case of REM reduces to the special case of the so called one-step RSB). 
Unfortunately, this is not more than 
a {\it heuristic} procedure which at present has no rigorous mathematical grounding.
On the other hand, it should be noted that 
during last decade remarkable progress has been 
achieved in mathematically rigorous derivations of various results previously obtained in
terms of the replica method. 
A number of rigorous results have been obtained which prove the validity of the cavity method 
for the entire class of the random satisfiability problems
revealing the physical phenomena similar to
what happens in REM and which are described by the one-step RSB solution
(see e.g. \cite{Bayati} and references there in).
The results obtained in terms of the continuous RSB scheme developed for
mean-field spin glasses has been also confirmed by independent
mathematically rigorous calculations \cite{guerra}.

Recently a notable progress has been achieved in the replica calculation technique itself
\cite{LeDoussal,Dotsenko}. This technique  does not require performing
analytic continuation to non-integer values of the replica parameter $n$, and 
formally makes possible to compute an entire free energy distribution function
summing over {\it all} integer momenta $Z(n)$.
In this brief communication I would like to present very simple 
replica calculations which do not involve such tricks like  RSB "magic operations", 
and which, nevertheless, recover well known result
for the average free energy of REM at all temperatures including the phase transition
into the low-temperature phase (which is usually called the one-step RSB state).


\section{Replica technique}

 By definition the partition function $Z$ of a given sample  
is related to its free energy $F$ via
\begin{equation}
\label{2.1}
Z = \exp( -\beta  F)
\end{equation}
The free energy $F$  is defined for a specific 
realization of the disorder and thus represent a random variable. 
Taking the $n$-th power of both sides of this relation 
and performing the disorder averaging  we obtain
\begin{equation}
\label{2.2}
\overline{Z^{n}} \equiv Z(n) = \overline{\exp( -\beta n F) }
\end{equation}
where the quantity in the lhs of the above equation is called the 
{\it replica partition function}.
The averaging in the rhs of the above equation can be represented in terms of the 
free energy distribution function $P(F)$. 
In this way we arrive to the following general relation 
between the replica partition function $Z(n)$ and the distribution function 
of the free energy  $P(F)$:
\begin{equation}
\label{2.3}
   Z(n) \; = \; \int_{-\infty}^{+\infty} dF \, P (F) \;  \exp\bigl(-\beta n F\bigr)
\end{equation}
 The above equation is the bilateral Laplace transform of  the function $P(F)$,
and it looks as if, at least formally, it allows to restore this function via inverse Laplace transform 
 of the replica partition function $Z(n)$:
\begin{equation}
\label{2.4}
 P(F) \; = \; \int_{-i\infty}^{+i\infty} \frac{d(n\beta)}{2\pi i} 
                              Z(n) \; \exp( \beta n \, F)
\end{equation}
In order to do so, first one has to compute 
$Z(n)$ for an  {\it arbitrary }
integer $n$ and then perform  analytical continuation of this function 
from integer to arbitrary complex values of $n$. This is the standard  strategy of the replica method
in disordered systems where it is well known that  very often the procedure of such analytic
continuation turns out to be rather controversial point  \cite{zirnbauer1,REM}. 

Usually the free energy of a given random system is expected to be the extensive quantity:
 $F = V f$, where $V$ is the volume of the system and $f$ is (random) free energy
density described by some distribution function ${\cal P}_{V}(f)$ (which in general depends
on the volume $V$). Substituting this into eq.(\ref{2.4}) and introducing a new integration
parameter $s = \beta n V$ we get
\begin{equation}
\label{2.5}
 {\cal P}_{V}(f) \; = \; \int_{-i\infty}^{+i\infty} \frac{ds}{2\pi i} 
                              {\cal Z}_{V}(s) \; \exp( s f)
\end{equation}
where ${\cal P}_{V}(f) = V P(Vf)$ and ${\cal Z}_{V}(s) = Z(s/\beta V)$.
If in the thermodynamic limit $V\to \infty$ the phenomenon of {\it selfaveraging}
takes place the limiting free energy distribution function becomes $\delta$-like:
$\lim_{V\to\infty} {\cal P}_{V}(f) = \delta\bigl(f - f(\beta)\bigr)$ where $f(\beta)$ is
the mean free energy density which is the quantity of the first interest in the disordered systems. 
According to the above equation, this means that 
the limiting replica partition function is expected to take the form
 $\lim_{V\to\infty} {\cal Z}_{V}(s) = \exp\bigl(-s f(\beta)\bigr)$ where the parameter $s$ 
remains {\it finite}. Since, by its definition, $s = \beta n V$, this implies that in the limit $V\to \infty$,
the replica parameter must go to zero, $n \sim 1/V \to 0$. 

The problem is that before taking the thermodynamic limit, the replica partition function ${\cal Z}_{V}(s)$ 
has to be computed for finite volume $V$. It is well known that in many cases 
the finite size distribution functions
of random quantities are extremely singular objects, and only in the thermodynamic limit they converge
to smooth and nice shape. The typical example is provided by the eigenvalues distribution functions
in the random matrix theory (see e.g.\cite{Mehta}).
 For that reason it could be easier,
instead of the distribution function itself, to study its integral representation,
namely,
\begin{equation}
 \label{2.6}
W(x) \; = \; \int_{x}^{\infty} \; df \; {\cal P}(f)  
\end{equation}
By definition, the function $W(x)$  
gives the probability that the random quantity $f$ is bigger than a given value $x$.
It is clear that this function is much more "smooth" object that the distribution function itself:
even in the case that the finite system size function  ${\cal P}_{V}(f)$ represents a set
 delta-functions, its integral representation would be only a kind of step-like continuous curve.

Formally the thermodynamic limit probability function $W(x)$ can be defined as follows:
\begin{eqnarray}
 \label{2.7}
W(x) &=& \lim_{V\to\infty} \sum_{n=0}^{\infty} \frac{(-1)^{n}}{n!} 
\exp(\beta n V x) \; \overline{Z^{n}}
\\
\nonumber 
&=& \lim_{V\to\infty} \sum_{n=0}^{\infty} \frac{(-1)^{n}}{n!} 
\; \overline{\exp(\beta n V x - \beta n V f)}
\\
\nonumber 
& = &  \lim_{V\to\infty} \overline{\exp\bigl[-\exp\bigl(\beta V (x-f)\bigr)\bigr]} 
\\
\nonumber 
& = &  
\overline{\theta(f-x)}
\end{eqnarray}
which coincides with the definition, eq.(\ref{2.6}). 
Thus, according to eq.(\ref{2.7}), the probability function $W(x)$ 
can be computed in terms of the above replica partition function
$Z(n)$ by summing over all replica {\it integers}
\begin{equation}
 \label{2.8}
W(x) \; = \; \lim_{V\to\infty} \sum_{n=0}^{\infty} \frac{(-1)^{n}}{n!} 
\exp(\beta n V x) \; Z(n)
\end{equation}
Of course, in the case $Z(n) \sim \exp(n^2)$ at large $n$, (which is the case of REM)
the above series is not that innocent. Here in accordance with the {\it troubles conservation law}
instead of the problem of analytic continuation to non-integer $n$'s we are facing formally divergent series.
Nevertheless, it can be shown that  such {\it sign alternating} series  can be easily 
regularized, similarly to formally divergent sign alternating series
$\sum_{k=0}^{\infty} (-1)^{k} a^{k} \; = \; (1+a)^{(-1)}$ which at $|a| > 1$ is well defined as
the analytic continuation from the region $|a| < 1$. 

\section{Random Energy Model}

\subsection{Definition}

The Random Energy Model is defined as a set of $M = 2^{N}$ states, characterized by
random energies $\{ E_{i}\} \; (i = 1, 2, ...,M)$ which are considered as independent
quenched random parameters described by the Gaussian distribution 
\begin{equation}
 \label{3.1}
{\cal P}[E_{1}, E_{2}, ...., E_{M}] \; = \; 
\prod_{i=1}^{M} \Biggl[\frac{1}{\sqrt{2\pi N}} \; \exp\Bigl(-\frac{E_{i}^{2}}{2N}\Bigr) \Biggr]
\end{equation}
Correspondingly, the partition function of REM is
\begin{equation}
 \label{3.2}
Z \; = \; \sum_{i=1}^{M} \; \exp(-\beta E_{i})
\end{equation}
which is a random quantity depending on $M$ random parameters $E_{1}, E_{2}, ..., E_{M}$.
The choice for the value  of $M = 2^{N}$ is motivated by
the fact that the free energy of this system (as will be shown below) is extensive in
$\ln M \propto N$, and thus it is the parameter $N$ which plays the role of the effective
"size" of the system (which is taken to infinity in the thermodynamic limit).
The particular form $2^{N}$ (instead of say, $\exp(N)$) is motivated by the idea to
imitate a random Ising system consisting of $N$ spins (having $2^{N}$ energy states, which,
of course,  are {\it not} independent).

\subsection{"Naive" solution}

Naively, one could propose very simple derivation for the average value of the free
energy  of this system. Since the partition function, eq.(\ref{3.2}), is given by the sum of
large number $M$ of independent random terms,  it could be approximated as follows:
\begin{equation}
 \label{3.3}
Z \; \simeq \; M \times \overline{\exp\bigl(-\beta E\bigr) } 
\end{equation}
Performing simple Gaussian averaging and substituting $M = 2^{N}$ we get 
\begin{equation}
 \label{3.4}
Z \; \simeq \;    2^{N} \, \exp\Bigl(\frac{1}{2} N \beta^{2} \Bigr) \; = \; 
                   \exp\Bigl(-\beta N f(\beta)\Bigr)
\end{equation}
where
\begin{equation}
 \label{3.5} 
f(\beta) \; = \; -\frac{1}{2}\beta - \frac{1}{\beta} \ln 2
\end{equation}
is the free energy density of the system. Correspondingly, for the entropy we get
\begin{equation}
 \label{3.6} 
S(T) \; = \; \beta^{2}\frac{d}{d\beta} f(\beta) \; = \; -\frac{1}{2} \beta^{2}  +  \ln 2
\end{equation}
Since we are dealing with the discrete system, one can immediately note that something is 
very wrong, as the entropy becomes negative  for $\beta > \sqrt{2\ln2}$. 
In fact, it turns out that there is the phase transition in the considered system
at $\beta_{c} = \sqrt{2\ln2}$, such that at $\beta > \beta_{c}$ (in the low temperature phase)
the system occupies only {\it finite number} of the lowest energy states. For that reason
the original hypothesis of the above derivation, that partition function, eq.(\ref{3.2}), 
contains macroscopic number of random terms, turns out to be wrong at low enough temperatures. 

\subsection{Rigorous solution}

The result of the rigorous (non-replica) derivation of the average free energy density
of REM \cite{REM} is in the following:
\begin{equation}
 \label{3.7} 
f(\beta) \; = \; \left\{\begin{array}{ll}
                    -\frac{1}{2}\beta - \frac{1}{\beta} \ln 2 \; , \; \; \mbox{at $\beta \leq \beta_{c} = \sqrt{2\ln 2}$}
                    \\
                    \\
                    - \sqrt{2\ln 2} \; , \; \; \; \; \; \;  \;\mbox{at $\beta \geq \beta_{c} $}
                   \end{array}
                   \right.
\end{equation}
Correspondingly, for the entropy density one gets:
\begin{equation}
 \label{3.8} 
S(\beta) \; = \; \left\{\begin{array}{ll}
                    -\frac{1}{2}\beta^{2} + \ln 2 \; , \; \; \mbox{at $\beta \leq \beta_{c}$}
                    \\
                    \\
                    0 \; , \; \; \; \; \; \; \; \; \; \; \; \;  \; \; \; \; \; \;\mbox{at $\beta \geq \beta_{c} $}
                   \end{array}
                   \right.
\end{equation}
The above  result for the entropy demonstrates that indeed in the low temperature phase 
the system effectively occupies only finite number of the lowest energy states.

\subsection{Replica approach}

In terms of the replica approach for the $n$-th momentum of the partition function, eq.(\ref{3.2}),
we get

\begin{eqnarray}
 \nonumber
Z(n) & = & 
\overline{\Biggl[\sum_{i=1}^{M} \; \exp(-\beta E_{i})\Biggr]^{n} } 
\; = \; 
\sum_{m_{1},...m_{M}=0}^{n} \frac{n!}{m_{1}!  ... m_{M}!} 
\overline{\Biggl[\exp\Bigl(-\beta \sum_{i=1}^{M} E_{i}m_{i} \Bigr)\Biggr]} \;
{\boldsymbol \delta}\Bigl(\sum_{i=1}^{M} m_{i}, \; n\Bigr)
\\
\nonumber
\\
& = &
\sum_{m_{1},...m_{M}=0}^{n} \frac{n!}{m_{1}!  ... m_{M}!} 
\exp\Bigl(\frac{1}{2} N \beta^{2} \sum_{i=1}^{M} m_{i}^{2} \Bigr) \;
{\boldsymbol \delta}\Bigl(\sum_{i=1}^{M} m_{i}, \; n\Bigr)
\label{3.9}
\end{eqnarray}
where ${\boldsymbol \delta}\bigl(p, q\bigr)$ is the kronecker symbol. 
Reorganizing the terms, we can also get the expression which would contain
summations over only non-zero values of $m$'s:
\begin{equation}
 \label{3.11}
Z(n) \; = \; 
\sum_{k=1}^{M}  \frac{M!}{k! (M-k)!} 
                            \sum_{m_{1}, ..., m_{k}=1}^{n} 
                            \frac{n!}{m_{1}!  ... m_{k}!} 
\exp\Bigl(\frac{1}{2} N \beta^{2} \sum_{\alpha=1}^{k} m_{\alpha}^{2} \Bigr) \;
{\boldsymbol \delta}\Bigl(\sum_{\alpha=1}^{k} m_{\alpha}, \; n\Bigr)
\end{equation}
We see that at large $n$,
\begin{equation}
 \label{3.12} 
Z(n\gg 1) \sim \exp\bigl[ C n^{2}\bigr]
\end{equation}
which indicates that analytic continuation of the replica partition function $Z(n)$ 
for non-integer $n$ is ambiguous \cite{REM}. 


\subsection{"RSB-magic" solution}

Heuristic RSB procedure for computing the thermodynamic limit average free energy using the 
above expression for the replica partition function, eq.(\ref{3.11}), is in the following.

(1) All $m_{\alpha}$'s are taken to be equal: $\; m_{1} = ... = m_{k} = m$. Correspondingly,
the constrain  $\sum_{\alpha=1}^{k} m_{\alpha} = n $ turns into $k m = n$ which fixes $k = n/m$ 
In this way the expression (\ref{3.11}) will contain no summations any more. 
Since $k\ll M$ we can estimate $M!/(M-k)! \; \sim \; M^{k} = \exp\bigl(k N \ln 2)$.
Then, neglecting all pre-exponential
factors (which are not extensive in $N$) the expression in eq.(\ref{3.11}) is estimated as follows:
\begin{equation}
 \label{3.13} 
 Z(n) \; \sim \; \exp\Bigl(\frac{1}{2} N \beta^{2} n m \; + \frac{Nn}{m} \ln 2\Bigr)
\; = \; \exp\Bigl(-\beta N n \, f(\beta, m)\Bigr) 
\end{equation}
where
\begin{equation}
 \label{3.14} 
 f(\beta,m) \; = \; -\frac{1}{2} \beta m  - \frac{1}{\beta m} \ln 2
\end{equation}
is a kind of the "trial free energy". The  expression for $f(\beta, m)$ explicitly depends on $m$ 
which is considered as an {\it optimization parameter}.

(2) According to its definition the parameter $m$ is constrained by 
the condition $1 \leq m \leq n$ (the fact that  it must be multiple of $n$ is  ignored).
In fact, this condition is reformulated in somewhat different way, namely: 
the value of the parameter $m$ must be {\it between} 1 and $n$.
Thus, after taking the limit $n\to 0$ the above constrain turns into $0 \leq m \leq 1$,
which implies that $m$ turns into a continuous real parameter.

(3) The parameter $m$ is defined by the condition that the "free energy"  $f(\beta, m)$, eq.(\ref{3.14}),
has a {\it maximum} at the interval  $0 \leq m \leq 1$. Thus, by definition, the physical
free energy of the system is 
\begin{equation}
 \label{3.15}
f(\beta) \; = \; \max_{0 \leq m \leq 1} \, f(\beta, m)
\end{equation}
One can easily check that at $\beta \leq \beta_{c} = \sqrt{2\ln 2}$ the maximum of the function  $f(\beta, m)$
in the interval $0 \leq m \leq 1$  is achieved at $m=1$, which  yields 
$f(\beta)  =  -\frac{1}{2}\beta - \frac{1}{\beta} \ln 2$. On the other hand, at 
$\beta \geq \beta_{c}$ the maximum of the function  $f(\beta, m)$
 is achieved at $m = \beta_{c}/\beta < 1$, which  yields 
$f(\beta)  =  - \sqrt{2\ln 2}$. Thus we see that the above "magic" manipulations
 nicely reproduce the rigorous solution, eq.(\ref{3.7}). All that is quite
impressive, but, of course, the above "derivation" can not be considered as grounded.

\subsection{Rigorous replica solution}

In fact rigorous replica calculations of the probability function $W(x)$, eq.(\ref{2.8}), 
 are rather simple. Substituting the expression for the replica partition function,
eq.(\ref{3.9}) into eq.(\ref{2.8}) we get
\begin{equation}
\label{3.16}
W(x) \; = \; \lim_{N\to\infty}  \sum_{n=0}^{\infty} \frac{(-1)^{n}}{n!} 
\sum_{m_{1},...m_{M}=0}^{\infty} \frac{n!}{m_{1}!  ... m_{M}!} 
\exp\Bigl(\frac{1}{2} N \beta^{2} \sum_{i=1}^{M} m_{i}^{2} + N\beta n x \Bigr) \;
{\boldsymbol \delta}\Bigl(\sum_{i=1}^{M} m_{i}, \; n\Bigr)
\end{equation}
where $M = 2^{N}$ and due to the presence of the kronecker symbol the summations over $m_{i}$'s can be
extended to infinity. Summing over $n$ we can lift the constrain $n = \sum_{i=1}^{M} m_{i}$
which provide independent summations over  $m_{i}$'s:
\begin{equation}
\label{3.18}
W(x) \; = \; \lim_{N\to\infty} \prod_{i=1}^{M} \Biggl(
\sum_{m_{i}=0}^{\infty}
\frac{(-1)^{m_{i}}}{m_{i}!}
\exp\Bigl[\frac{1}{2} N \beta^{2}  m_{i}^{2} + N \beta x m_{i} \Bigr] \Biggr)
\end{equation}
Elementary summations of these series yields:
\begin{equation}
\label{3.19}
W(x) \; = \; 
\lim_{N\to\infty} \Bigl[G(N,x)\Bigr]^{2^{N}}
\end{equation}
 where
\begin{equation}
 \label{3.20} 
 G(N,x) \; = \; \sqrt{\frac{N}{2\pi}} \int_{-\infty}^{+\infty} d\xi \; 
\exp\Biggl(-\frac{1}{2} N \xi^{2} -\exp\bigl(\beta N (x-\xi)\bigr)\Biggr) 
\end{equation}
Taking the limit $N \to \infty$ one easily obtain the following result (see Appendix):

\vspace{5mm}

(1) In the region $\beta \leq \sqrt{2\ln 2} \equiv \beta_{c}$:

\begin{equation}
 \label{3.21} 
W(x) \; = \; \left\{\begin{array}{ll}
                    1 \; , \; \; \mbox{for $x < - \Bigl(\frac{1}{2}\beta +\frac{1}{\beta} \ln 2\Bigr)$}
                    \\
                    \\
                    0 \; , \; \; \mbox{for $x > - \Bigl(\frac{1}{2}\beta +\frac{1}{\beta} \ln 2\Bigr)$}
                   \end{array}
                   \right.
\end{equation}

(2) In the region $\beta \geq \sqrt{2\ln 2} $:

\begin{equation}
 \label{3.22} 
W(x) \; = \; \left\{\begin{array}{ll}
                    1 \; , \; \; \mbox{for $x < - \sqrt{2\ln 2}$}
                    \\
                    \\
                    0 \; , \; \; \mbox{for $x > - \sqrt{2\ln 2}$}
                   \end{array}
                   \right.
\end{equation}
According to the definition, eq.(\ref{2.6}), the above result means that 
in the thermodynamic limit the free energy distribution function of the considered model
is the delta-function (which means that the system is selfaveraging):
\begin{equation}
 \label{3.23} 
 P(f) \; = \; \delta\bigl(f - f(\beta)\bigr)
\end{equation}
where the  free energy $f(\beta)$ coincides with the one obtained in the rigorous
(non-replica) solution, eq.(\ref{3.7}).

\section{Conclusions}

In its traditional formulation the replica method procedure is in the following: 
first, one has to calculate the disorder average of the integer $n$-th power of the partition function,
$\overline{Z^{n}} = Z(n)$;
second, one has to perform an analytic continuation of this function  for arbitrary real 
or complex values of $n$; and third, one has to take the limit $n\to 0$ (to get the 
average free energy) or  integrate over complex $n$ (to derive the
free energy distribution function).
 The third step is usually accompanied by taking the thermodynamic limit,
which assumes that the system size $L$ is taken to infinity. 
The prescription of the replica method indicates that the two limits, $n\to 0$ and $L\to\infty$, 
has to be taken simultaneously such that the product
$n L^{\omega}$ (where an exponent $\omega$ defines the scaling of the free energy with the system size)
is kept finite.

In fact, the whole experience of the replica calculations in 
disordered systems shows that except for trivial cases this program,
as it is formulated above, is never followed (for more detailed discussion of this issue
see \cite{replicas}). The typical illustration
is provided by the studies of the Sherrington-Kirkpatrick model of spin glasses \cite{SK}. 
The replica solution of this model which is generally believed to be correct is 
derived in terms of the RSB technique \cite{RSB-general} in which  all the above three steps, 
(computing $Z(n)$, analytic continuation in $n$ and the limits $n\to 0$ and $L\to\infty$) 
are performed simultaneously. 

In this paper an alternative replica technique has been discussed. In terms of this approach
no analytic continuation for non-integer values of the replica parameter $n$ is required, and instead 
the summation over all positive integer momenta of the partition function has to be performed.
Earlier this method has been successfully applied for solving the one-dimensional  directed polymer problem
\cite{LeDoussal,Dotsenko} 
In this paper it has been demonstrated that in terms of this technique  the rigorous solution of the
Random Energy Model takes just a few lines. Of course, real challenge would be to find an alternative solution
for the SK model. Unfortunately, here the situation is much more complicated, as already at the stage of calculation
of the replica partition function $Z(n)$ (for integer $n$) the saddle point approximation 
is required which doesn't seem to be legitimate in terms of the present technique. In any case further
systematic studies of the considered approach is required. 


\acknowledgments

I am grateful to Sergei Nechaev, Bernard Derrida and Herbert Spohn for fruitful
discussions of this work.

This work was supported in part by the IRSES grant DCP Phys Bio.

\vspace{10mm}

\begin{center}

\appendix{\Large \bf Appendix }

\end{center}

\newcounter{A}
\setcounter{equation}{0}
\renewcommand{\theequation}{A.\arabic{equation}}

\vspace{5mm}

Let us study the properties of the function, eq.(\ref{3.20}),
\begin{equation}
 \label{A1} 
 G(N,x) \; = \; \sqrt{\frac{N}{2\pi}} \int_{-\infty}^{+\infty} d\xi \; 
\exp\Biggl(-\frac{1}{2} N \xi^{2} -\exp\bigl(\beta N (x-\xi)\bigr)\Biggr) 
\end{equation}
in the limit of large $N$. First of all one can easily see that at $x > 0$ and $N \gg 1$,
\begin{equation}
 \label{A2} 
 G(N,x) \; \sim \; \exp\Bigl(-\frac{1}{2} N x^{2} \Bigr) \; \to 0
\end{equation}
Substituting this into eq.(\ref{3.19}), we find that
\begin{equation}
 \label{A3} 
W(x>0) \; = \; 0
\end{equation}

At $x < 0$ we can represent the function, eq.(\ref{A1}), as the sum of two 
contributions:
\begin{equation}
 \label{A4} 
 G(N,x) \; = \; G_{1}(N,x) + G_{2}(N,x)
\end{equation}
where
\begin{equation}
 \label{A5} 
 G_{1}(N,x) \; = \; \sqrt{\frac{N}{2\pi}} \int_{-\infty}^{-|x|} d\xi \; 
\exp\Biggl(-\frac{1}{2} N \xi^{2} -\exp\bigl(-\beta N (|x|+\xi)\bigr)\Biggr) 
\end{equation}
and 
\begin{equation}
 \label{A6} 
 G_{2}(N,x) \; = \; \sqrt{\frac{N}{2\pi}} \int_{-|x|}^{+\infty} d\xi \; 
\exp\Biggl(-\frac{1}{2} N \xi^{2} -\exp\bigl(-\beta N (|x|+\xi)\bigr)\Biggr) 
\end{equation}
Simple analysis shows that
\begin{equation}
 \label{A7} 
 G_{1}(N,x)\Big|_{N\gg1} \; \sim \; \exp\Bigl(-\frac{1}{2} N x^{2}\Bigr)
\end{equation}
The function $G_{2}(N,x)$ in the limit of large $N$ can be estimated as follows:
\begin{eqnarray}
\nonumber 
 G_{2}(N,x) & \simeq & \sqrt{\frac{N}{2\pi}} \int_{-|x|}^{+\infty} d\xi \; 
\exp\Bigl(-\frac{1}{2} N \xi^{2}\Bigr) \Biggl[1 -\exp\bigl(-\beta N (|x|+\xi)\bigr)\Biggr] 
\\
\nonumber
\\
 & \simeq & 1 \; - \; \sqrt{\frac{N}{2\pi}} \int_{-|x|}^{+\infty} d\xi \; 
\exp\Bigl(-\frac{1}{2} N \xi^{2} -\beta N \xi -\beta N |x| \Bigr)
 \label{A8}
\end{eqnarray}
The function $\varphi(\xi) = -\frac{1}{2} N \xi^{2} - \beta \xi$ has the maximum at
$\xi_{*} = - \beta$. Thus
\begin{equation}
 \label{A9}
\sqrt{\frac{N}{2\pi}} \int_{-|x|}^{+\infty} d\xi \; 
\exp\Bigl(-\frac{1}{2} N \xi^{2} -\beta N \xi -\beta N |x| \Bigr)
\; \sim \; 
\left\{\begin{array}{ll}
                     \exp\Bigl(-\frac{1}{2} N x^{2} \Bigr)\; , 
                      \; \;  \; \;  \; \;  \; \; \;  \; \; \;  \; \; \mbox{for $|x| < \beta$}
                    \\
                    \\
                    \exp\Bigl[-\beta N\bigl(|x|-\frac{1}{2}\beta\bigr)\Bigr] \; , \; \; \mbox{for $|x| > \beta$}
                   \end{array}
                   \right.
\end{equation}
The results, eqs.(\ref{A7}), (\ref{A8}) and (\ref{A9}), demonstrate that at $N \gg 1$ and negative $x$
the function $G(N,x)$, eq.(\ref{A1}), takes the form
\begin{equation}
 \label{A10}
G(N,x) \; \simeq \; 1  -  g(N,x)
\end{equation}
where
\begin{equation}
 \label{A11}
g(N,x) 
\; \sim \; 
\left\{\begin{array}{ll}
                     \exp\Bigl(-\frac{1}{2} N x^{2} \Bigr)\; , 
                      \; \;  \; \;  \; \;  \; \; \;  \; \; \;  \; \; \mbox{for $|x| < \beta$}
                    \\
                    \\
                    \exp\Bigl[-\beta N\bigl(|x|-\frac{1}{2}\beta\bigr)\Bigr] \; , \; \; \mbox{for $|x| > \beta$}
                   \end{array}
                   \right.
\end{equation}
Substituting eqs.(\ref{A10}) and (\ref{A11}) into eq.(\ref{3.19}) we get
\begin{equation}
\label{A12}
W(x) \; = \; 
\lim_{N\to\infty} \; \exp\Bigl[-\psi(N,x)\Bigr]
\end{equation}
 where
\begin{equation}
 \label{A13}
\psi(N,x) \equiv g(N,x) \; 2^{N} 
\; \sim \; 
\left\{\begin{array}{ll}
                     \exp\Bigl(-\frac{1}{2} N x^{2} + N \ln 2\Bigr)\; , 
                      \; \;  \; \;  \; \;  \; \; \;  \; \; \;  \; \; \mbox{for $|x| < \beta$}
                    \\
                    \\
                    \exp\Bigl[-\beta N\Bigl(|x|-\frac{1}{2}\beta -\frac{1}{\beta}\ln 2  \Bigr)\Bigr] \; , 
                      \; \; \mbox{for $|x| > \beta$}
                   \end{array}
                   \right.
\end{equation}
Simple analysis of this expression yields:

\vspace{3mm}

(a) at $|x| < \beta$,

\begin{equation}
 \label{A14}
\lim_{N\to\infty} \psi(N,x) \; = \; 
\left\{\begin{array}{ll}
                     0       \; , \; \;  \; \; \;  \; \;  \; \, \mbox{for $|x| > \sqrt{2\ln 2}$}
                    \\
                    \\
                    +\infty  \; , \; \;  \; \; \mbox{for $|x| < \sqrt{2\ln 2}$}
                   \end{array}
                   \right.
\end{equation}

(b) at $|x| > \beta$,

\begin{equation}
 \label{A15}
\lim_{N\to\infty} \psi(N,x) \; = \; 
\left\{\begin{array}{ll}
                     0       \; , 
\; \;  \; \; \;  \; \;  \; \, \mbox{for $|x| > \frac{1}{2}\beta + \frac{1}{\beta} \ln 2$}
                    \\
                    \\
                    +\infty  \; , 
                 \; \;  \; \; \mbox{for $|x| < \frac{1}{2}\beta + \frac{1}{\beta} \ln 2$}
                   \end{array}
                   \right.
\end{equation}

Substituting eqs.(\ref{A14}) and (\ref{A15}) into eq.(\ref{A12}) we find:

\vspace{3mm}

(a) at $|x| < \beta$,

\begin{equation}
 \label{A16}
W(x) \; = \; 
\left\{\begin{array}{ll}
                     1  \; , \; \; \; \; \mbox{for $|x| > \sqrt{2\ln 2}$}
                    \\
                    \\
                     0  \; , \; \; \; \; \mbox{for $|x| < \sqrt{2\ln 2}$}
                   \end{array}
                   \right.
\end{equation}

(b) at $|x| > \beta$,

\begin{equation}
 \label{A17}
W(x) \; = \; 
\left\{\begin{array}{ll}
                     1  \; , \; \; \; \;  \mbox{for $|x| > \frac{1}{2}\beta + \frac{1}{\beta} \ln 2$}
                    \\
                    \\
                     0  \; , \; \; \; \; \mbox{for $|x| < \frac{1}{2}\beta + \frac{1}{\beta} \ln 2$}
                   \end{array}
                   \right.
\end{equation}

One can easily see that the above results, eq.(\ref{A16}) (valid for $|x| < \beta$) , and 
eq.(\ref{A17}) (valid for $|x| > \beta$) are equivalent to eqs.(\ref{3.21})-(\ref{3.22}).

\end{document}